\documentclass[preprint]{revtex4}
\usepackage{bm}
\usepackage{verbatim}
\usepackage{epsfig}

\begin{document}
\title{Vortex Properties of a Resonant Superfluid}
\author{Meng Gao}
\author{Hongyu Wu}
\author{Lan Yin}
\email{yinlan@pku.edu.cn} \affiliation{School of Physics, Peking
University, Beijing 100871, P. R. China}
\date{May 8, 2006}

\begin{abstract}
The properties of a vortex in a rotating superfluid Fermi gas are
studied in the unitary limit.  A phenomenological approach based on
Ginzburg-Landau theory is developed for this purpose.  The density
profiles, including those of the normal fluid and superfluid, are
obtained at various temperatures and rotation frequencies.  The
superfluid and normal fluid densities can be identified from the
angular momentum density.  The total free energy and angular
momentum of the vortex are also obtained.

\end{abstract}


\maketitle

\section{Introduction}
The field of BEC-BCS crossover has attracted great attention from
both experimental and theoretical sides.  In experiments, the system
is studied near Feshbach resonances so that the scattering length
can be tuned by an applied magnetic field.  At low temperatures,
when the scattering length is tuned from negative side to positive
side, the system evolves from a BCS state to a molecular BEC state.
The observation of molecular BEC was first reported in references
\cite{Jochim, Greiner}.  When the scattering length was fast tuned
from the negative side to the positive side of the resonance,
condensation of Fermion pairs was observed \cite{jin, ketterle04}.
Since then quite a few experiments have focused on various
properties of the system, including collective excitations
\cite{thomas, chin}, single-particle excitations \cite{chin2, jin2},
and thermodynamic properties \cite{thomas2}.  The observation of
vortex lattice \cite{ketterle05} has provided a conclusive evidence
of superfluidity across the resonance in this system.

The vortex structure in superfluid fermi gas have been studied
theoretically \cite{bruun1, torma, nygaard, nygaard2, bulgac,
tempere, levin, ho}.  However most of these works focus on the
non-rotating case, which is different from the experimental
situation. To consider the effect of rotation, we have developed a
phenomenological approach, based on our previous work on the
phase-slip phenomenon in the resonant Fermi superfluid \cite{yin}.

In this paper, we focus on the properties of a vortex in the unitary
limit \cite{Ho2}, because it is a special strong-interaction region
where the magnitude of the scattering length is much bigger than the
interparticle distance.  In the unitary limit, the interparticle
distance replaces the scattering length as the most important length
scale, and the physical properties become unitary. In the following,
we first present our phenomenological approach and then study the
vortex structure at various temperatures and rotation frequencies.

\section{Formalism}
  In our previous work \cite{yin}, we developed a phenomenological
theory for the resonant Fermi superfluid based on Ginzburg-Landau
theory. To consider the effect of rotation, we study the system in
the rotating frame, in which the single-particle Hamiltonian is
given by
\begin{equation}
H-\Omega L_{z}=\frac{1}{2m}(p_{\bot}^{2}+p_{z}^{2})
+\frac{1}{2}m(\omega^{2}_{\bot}r_{\bot}^{2}+ \omega_{z}z^{2})-
\Omega \mathbf{\hat{z}\cdot r_{\bot} \times p_{\bot}},
\end{equation}
where $\mathbf{p}_{\bot}=(p_{x}, p_{y},0)$, $\mathbf{r}_{\bot}=(x,
y,0)$, $\omega_{\bot}$ and $\omega_{z}$ are the transverse and
longitudinal trapping frequencies, and $\Omega$ is the rotation
frequency.  Here the trapping potential is assumed to have the
rotation symmetry in the transverse plane.  The single-particle
Hamiltonian can be rewritten as
\begin{equation}\label{eqrt}
H-\Omega L_{z}=\frac{1}{2m}(\mathbf{p}-{e \mathbf{A}\over c})^2
+\frac{1}{2}m[(\omega^2_{\bot}-\Omega^2)r_{\bot}^2+\omega_{z}z^2],
\end{equation}
where $\mathbf{p}-e\mathbf{A}/c$ is the canonical momentum, and
$e\mathbf{A}/c=m\Omega \hat{z}\times \mathbf{r}_{\bot}$.  From Eq.
(\ref{eqrt}), we conclude that the motion of a charge-neutral
fermion in a rotating frame is analogous to an electron moving in a
vector potential $\mathbf{A}$ and in a trapping potential reduced in
the transverse direction.

In the superfluid state, the free energy of a fermi gas is lower
than that in the normal state, and the energy difference is defined
as the condensation energy. The total free energy $F$ is the sum of
condensation energy $F_{C}$ and the free energy of the normal state
$F_{N}$,
\begin{equation}
 F=F_{C}+F_{N}.
\end{equation}
The free energy of the normal state $F_N$ is generally a complicated
function of temperature, particle density, trapping potential, the
scattering length, and the rotation frequency.  According to
Ginzburg-Landau theory, the condensation energy can be expressed as
a functional of the vector potential $\mathbf{A}$ and the order
parameter $\phi$ which is proportional to the energy gap of the
fermion excitation,
\begin{equation}\label{lc}
F_{C}=\int d^{3}r [-\frac{\hbar^2}{2m}\phi^{\ast}(\mathbf{r})
(\nabla-\frac{2ie\mathbf{A}}{\hbar c})^2\phi(\mathbf{r}) +
\alpha(\mathbf{r})|\phi(\mathbf{r})|^2+\frac{\beta(\mathbf{r})}{2}
|\phi(\mathbf{r})|^4],
\end{equation}
where $\alpha$ and $\beta$ are coefficients.  Although
Ginzburg-Landau theory was originally proposed for describing
superconducting system near the transition, it can often be applied
to most of the region below the transition temperature.

When the system is not rotating, $\Omega=0$, the sign change of the
coefficient $\alpha$ signals the superfluid transition. Below the
transition temperature $T_C$, $\alpha<0$, and the system has lower
free energy in the superfluid state than in the normal state. When
the system is rotating, the transition temperature becomes a
function of the rotation frequency $\Omega$. In the following, the
symbol $T_C$ is defined as the transition temperature at $\Omega=0$.

In the superfluid state, the total free energy $F$ is at a minimum
point as a functional of the particle density $n(\mathbf{r})$ and
the superfluid order parameter $\phi(\mathbf{r})$,
\begin{eqnarray}
{\delta F \over \delta \phi^{\ast}({\bf r})}&=& 0, \label{s1} \\
{\delta F \over \delta n({\bf r})}&=&0. \label{s2}
\end{eqnarray}
 In Eq. (\ref{s2}), the derivative is taken under the constraint of
constant total-particle number.  In principle, if the free energy
$F$ is known, the distributions of particle density and order
parameter in the superfluid state can be solved from the saddle
point equations, Eq. (\ref{s1}, \ref{s2}).

At low temperatures and low rotation frequencies, when the Fermi
energy is much bigger than rotation energy, $E_F \gg \hbar \Omega$,
the average kinetic energy per particle is proportional to
$n^{2/3}({\bf r})$ in the Thomas-Fermi approximation \cite{Ho3}.  In
the unitary limit, the interparticle distance replaces the
scattering length, and the average interaction energy is also
proportional to $n^{2/3}({\bf r})$.   Therefore the free energy of
the normal state is approximately given by
\begin{equation}
F_N \approx \int d^3r \left[ {\hbar^2 \over 2 m} c_N  n^{5 \over
3}({\bf r})+V'({\bf r}) n({\bf r}) \right ],
\end{equation}
where $V'({\bf r}) \equiv m [(\omega^2_{\bot}-\Omega^2)r_{\bot}^2 +
\omega_{z}z^2]/2$ is the trapping potential reduced due to rotation,
and $c_N$ is a dimensionless constant.  From the experiment
\cite{thomas04}, the constant $c_N$ is estimated to be about 4.3. At
low rotation frequencies, $E_F \gg \hbar \Omega$, the constant $c_N$
does not depend on the rotation frequency $\Omega$.  In general, the
normal-state free energy $F_N$ should be temperature dependent.
However, near the transition temperature which is much smaller than
the Fermi temperature \cite{chin2}, the normal-state free energy
$F_N$ is weakly dependent on the temperature and the coefficient
$c_N$ is approximately temperature independent.

In the unitary limit, the coefficients $\alpha$ and $\beta$ in Eq.
(\ref{lc}) can be expressed in terms of the particle density $n$ and
temperature $T$. The coefficient $\alpha$ is approximately given by
$\alpha=c_{\alpha} k_B (T-T_C)$, where $c_{\alpha}$ is a
dimensionless constant and $k_B$ is the Boltzmann constant. The
coefficient $\beta$ is a function of the fermion density,
$\beta=c_{\beta} \hbar^2/[2 m n^{1/3}]$, where $c_{\beta}$ is a
dimensionless constant. The transition temperature in the unitary
limit is proportional to $n^{2/3}$, $k_B T_C=c_T \hbar^2 n^{2/3}/(2
m)$, where $c_T$ is a dimensionless constant.  In the experiment
\cite{chin2}, the transition temperature $T_C$ is about $25\%$ of
the Fermi temperature $T_F$, corresponding to $c_{T}=2.4$. Although
the zero-temperature properties of the unitary Fermi gas has been
numerically obtained \cite{Carlson}, the accurate values of the
constants $c_{\alpha}$ and $c_{\beta}$ near the transition
temperature are still unknown.  In the following, we use their
values in the weak-interaction case instead, $c_\alpha=1.47 c_T$ and
$c_\beta=2.94 c_T^2$ \cite{Popov}, and discuss how these parameters
affect the system properties near the end.  In the unitary limit,
the saddle-point equations (\ref{s1}, \ref{s2}) can be further
written as
\begin{eqnarray}
-{\hbar^2 \over 2 m} (\nabla-\frac{2ie\mathbf{A}}{\hbar c})^2
\phi({\bf r})+\alpha({\bf r}) \phi({\bf r})+\beta({\bf r})
|\phi({\bf r})|^2 \phi({\bf r}) &=& 0, \label{s1'} \\
{5 \hbar^2 \over 6 m} c_N n^{2/3}({\bf r})+V'({\bf r})-\mu-{\hbar^2
\over 3 m} c_T c_\alpha {|\phi({\bf r})|^2 \over n^{1/3}({\bf
r})}-{\hbar^2 \over 12m} c_{\beta} {|\phi({\bf r})|^4 \over
n^{4/3}({\bf r})} &=& 0, \label{s2'}
\end{eqnarray}
where $\mu$ is the chemical potential.

The angular momentum can be obtained from the free energy,
$L_z=-\partial F/\partial \Omega$. In the unitary limit, it is given
by
\begin{eqnarray}
L_z &=& \int d^3 r [ m \Omega r_{\bot}^2 n({\bf r})+ 2 \phi^{\ast}
(\mathbf{r}) (\hat{z}\times \mathbf{r}_{\bot}) \cdot (-i \hbar
\nabla-\frac{2e\mathbf{A}}{c}) \phi(\mathbf{r})] \\ &=& \int d^3 r
\{ m \Omega r_{\bot}^2 [n({\bf r})-4 |\phi(\mathbf{r})|^2]-2 i \hbar
\phi^{\ast} (\mathbf{r}) {\partial \over \partial \theta}
\phi(\mathbf{r}) \}, \label{lz}
\end{eqnarray}
where $\theta$ is the azimuthal angle around $\hat{z}$-axis. When
there is no superfluid current, the phase of the order parameter is
uniform and the angular momentum totally comes from the normal
fluid,
\begin{equation}
L_z=\int d^3 r m \Omega r_{\bot}^2 [n({\bf r})-4
|\phi(\mathbf{r})|^2].
\end{equation}
Therefore we identify $n_{n}({\bf r})\equiv n({\bf r})-4 |\phi({\bf
r})|^2$ as the density of the normal fluid and $n_{s}({\bf r})\equiv
4 |\phi({\bf r})|^2$ as the superfluid density.  In rotating
systems, if the density of the angular momentum can be measured
experimentally, the superfluid and normal fluid densities can be
obtained from Eq. (\ref{lz}).

\section{Vortex structure and thermodynamical properties}
  In a single-vortex state, the order parameter takes the form
$\phi(\mathbf{r})=f(r_{\bot},z)e^{i\theta}$, where for convenience
$f$ can be chosen as a positive function, $f=|\phi|$. The order
parameter and particle density of the vortex state can be obtained
from the saddle-point equations (\ref{s1'}, \ref{s2'}) which can now
be further simplified,
\begin{eqnarray}\label{saddle1}
-\frac{\hbar^2}{2m}(\frac{1}{r_{\bot}} \frac{\partial f}{\partial
r_{\bot}}+ \frac{\partial^2 f}{\partial r_{\bot}^2}+
\frac{\partial^2 f}{\partial z^2}- \frac{f}{r_{\bot}^2}) +2
m\Omega^2 r_{\bot}^2 f-2 \hbar \Omega f + \alpha f + \beta f^3 = 0,
\\ \label{saddle2} \frac{5\hbar^2}{6m}c_{N}n^{2/3}+V'-\mu
-\frac{\hbar^2}{3m}c_{T} c_{\alpha} \frac{f^2}{n^{1/3}}-
\frac{\hbar^2}{12m} c_{\beta} \frac{f^4}{n^{4/3}}=0.
\end{eqnarray}
The angular momentum in the vortex state can be obtained from Eq.
(\ref{lz}),
\begin{equation} \label{lz'}
L_z=\int d^3 r [ m \Omega r_{\bot}^2 (n-4 f^2)+2 \hbar f^2].
\end{equation}
From Eq. (\ref{lz'}), we can see that in a single-vortex state each
superfluid particle contributes $\hbar/2$ to the total angular
momentum.

  To solve the nonlinear differential equations Eq. (\ref{saddle1}) and
(\ref{saddle2}), it is necessary to use the numerical method. Here
we take the experimental parameters from Ref. \cite{ketterle05}, in
which the average number of Li$^6$ atoms is about two million, $N
\approx 2\times10^6$, the axial and radial trap frequencies are
given by $\nu_{z}=23$Hz and $\nu_{r}=57$Hz.  The nonlinear partial
differential equations (\ref{saddle1}) and (\ref{saddle2}) are
solved by the iteration method.

\begin{figure}
\centering \epsfig{file=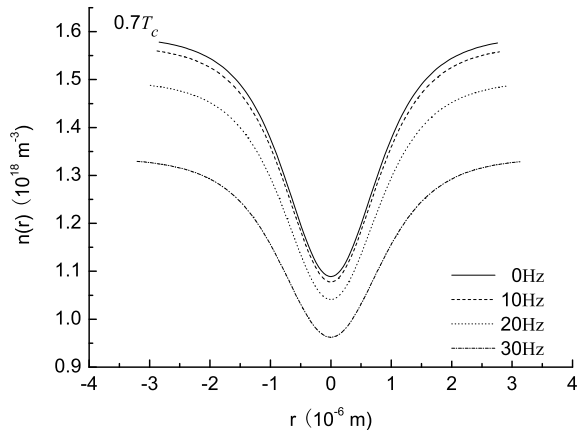, height=2.6in} \caption{The density
profile of the vortex at $T=0.7T_{C}$ and $z=0$.  The solid, dashed,
dotted, and dotted dash lines are density profiles at rotating
frequency $0$Hz, $10$Hz, $20$Hz, and $30$Hz.} \label{totaldensity-w}
\end{figure}
\begin{figure} \centering \epsfig{file=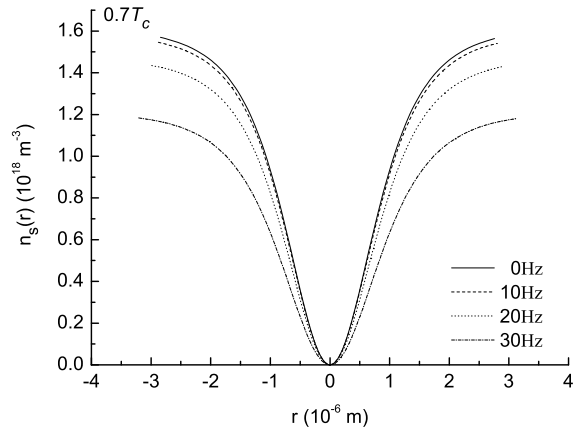, height=2.6in}
\caption{The superfluid density profile at $T=0.7T_{C}$ and $z=0$.
The solid, dashed, dotted, and dotted dash lines are superfluid
density profiles at rotating frequency $0$Hz, $10$Hz, $20$Hz, and
$30$Hz.} \label{superfluiddensity-w}
\end{figure}

In Fig. (\ref{totaldensity-w}), the density profiles of the vortex
are plotted at $T=0.7T_{C}$ and different rotation frequencies.  The
vortex produces a density dip at the center and the normal fluid is
dominant inside the vortex core.  This is because the superfluid
density vanishes at the center, as shown in Fig.
(\ref{superfluiddensity-w}).
At very low rotation frequencies, from Eq. (\ref{saddle1}) the size
of the vortex core is given by $\xi=\hbar/\sqrt{-2m\alpha(0)}$. When
the rotation frequency increases, the density decreases near the
center of the trap as shown in Fig. (\ref{totaldensity-w}), due to
the increase of the centrifugal force.  When the density decreases,
the coefficients $\alpha$ and $\beta$ in the Ginzburg-Landau theory
increase, and from Eq. (\ref{saddle1}) the size of the order
parameter decreases. As a result, the superfluid density also
decreases with the increase of the rotation frequency, and the size
of the vortex core increases, as shown in Fig.
(\ref{superfluiddensity-w}).

\begin{figure}
\centering \epsfig{file=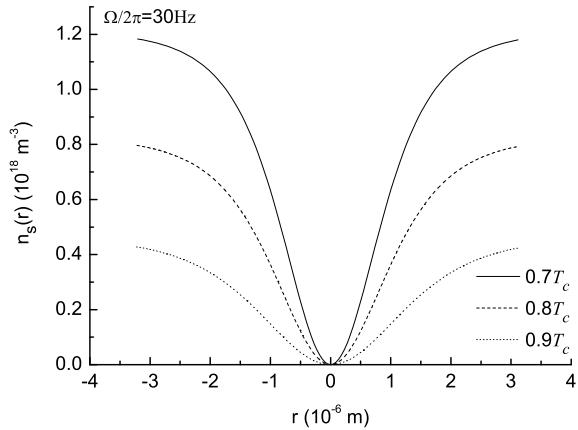, height=2.6in} \caption{The
superfluid density profile at $\Omega/(2\pi)=30$Hz and $z=0$. The
solid, dashed, and dotted lines are superfluid density profiles at
$0.7T_C$, $0.8T_C$, and $0.9T_C$.} \label{nstp}
\end{figure}

\begin{figure}
\centering \epsfig{file=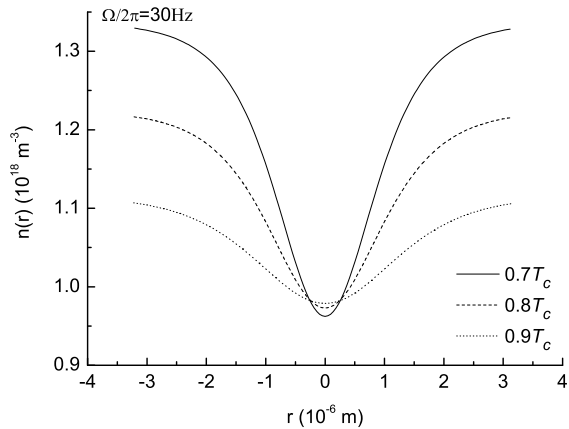, height=2.6in} \caption{The density
profile of the vortex at $\Omega/(2\pi)=30$Hz and $z=0$. The solid,
dashed, and dotted lines are the density profiles at $0.7T_C$,
$0.8T_C$, and $0.9T_C$.} \label{ntp}
\end{figure}

When the temperature increases, the superfluid density decreases as
shown in Fig. (\ref{nstp}).
The size of the vortex core becomes larger because the coefficient
$\alpha$ increases with the increase of temperature. The density
close to the vortex core also decreases as shown in Fig.
(\ref{ntp}).
To keep the total particle number constant, the densities at other
places increase.  As a result, the density at the center slightly
increases with the increase of temperature.

\begin{figure} \centering
\epsfig{file=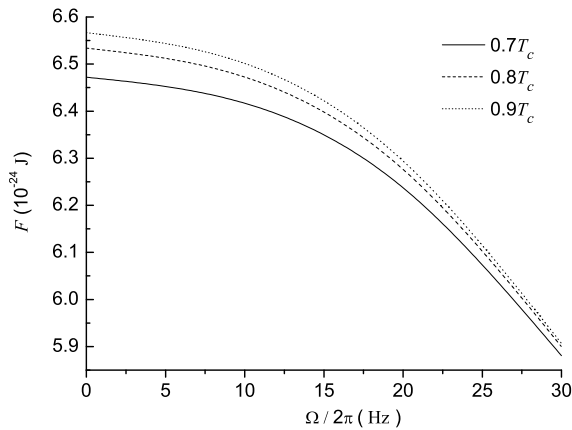, height=2.6in} \caption{The total free energy as
a function of rotation frequency. The solid, dashed, and dotted
lines are free energies at $0.7T_C$, $0.8T_C$, and $0.9T_C$.}
\label{lp}
\end{figure}

\begin{figure}
\centering \epsfig{file=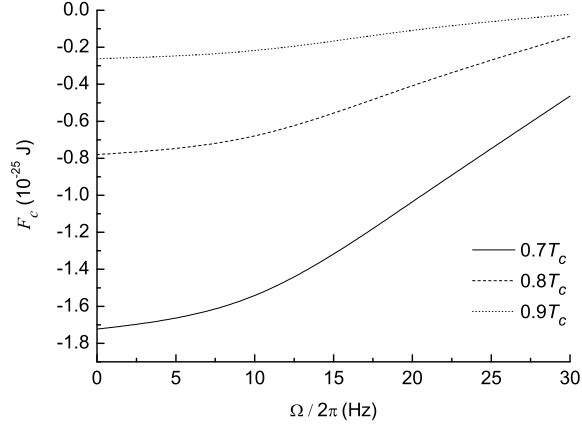, height=2.6in} \caption{The
condensation energy as a function of rotation frequency. The solid,
dashed, and dotted lines are condensation energies at $0.7T_C$,
$0.8T_C$, and $0.9T_C$.} \label{lcp}
\end{figure}

\begin{figure}
\centering \epsfig{file=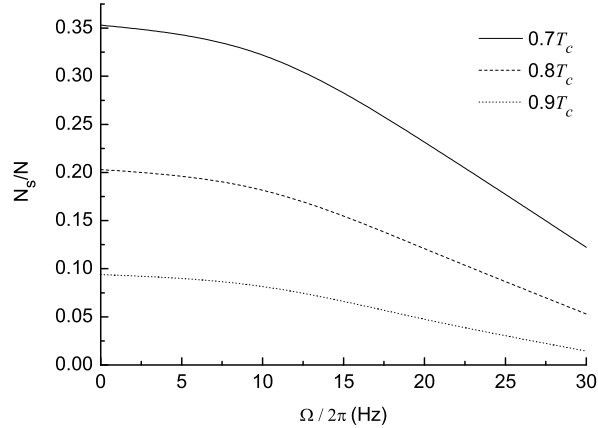, height=2.6in} \caption{The
superfluid fraction as a function of rotation frequency. The solid,
dashed, and dotted lines are the ratios of the superfluid atom
number to total atom number at $0.7T_C$, $0.8T_C$, and $0.9T_C$.}
\label{Nsp}
\end{figure}

\begin{figure}
\centering \epsfig{file=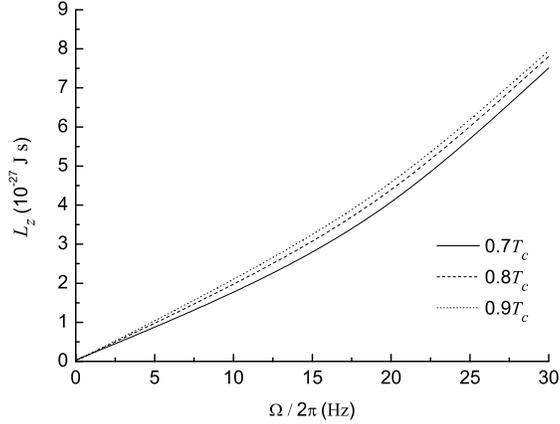, height=2.6in} \caption{The total
angular momentum as a function of rotation frequency. The solid,
dashed, and dotted lines are the angular momentum at $0.7T_C$,
$0.8T_C$, and $0.9T_C$.} \label{mp}
\end{figure}

The total free energy as a function of the rotation frequency is
plotted in Fig. (\ref{lp}).
As the rotation frequency increases, the free energy decreases
mainly due to the reduction of the trapping potential.  As the
temperature increases, the free energy also increases. In contrast,
the absolute value of the condensation energy decreases with the
increase of temperature as shown in Fig. (\ref{lcp}).
The absolute value of the condensation energy also decreases with
the increase of rotation frequency due to the decrease of the total
superfluid fraction as shown in Fig. (\ref{Nsp}).
The total angular momentum almost increases linearly with the
rotation frequency as shown in Fig. (\ref{mp}).
The linear dependence comes from the normal fluid contribution, and
in contrast the superfluid in the vortex state contributes a small
constant to the total angular momentum.

\section{Discussion and Conclusion}
\begin{figure}
\centering \epsfig{file=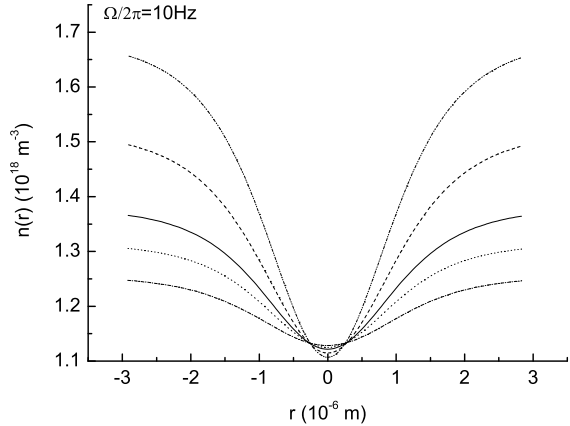, height=2.6in} \caption{The density
profile at $T=0.9T_C$ and $\Omega/(2\pi)=10$Hz.  The solid line is
the density profile with the weak-interaction expressions of
$c_\alpha$ and $c_\beta$.   The dotted-dash and double-dotted-dashed
lines are the density profiles with $c_\alpha$ decreased and
increased by $20\%$ respectively.  The dashed and dotted lines are
the density profiles with $c_\beta$ decreased and increased by
$20\%$ respectively. } \label{n10p}
\end{figure}

\begin{figure}
\centering \epsfig{file=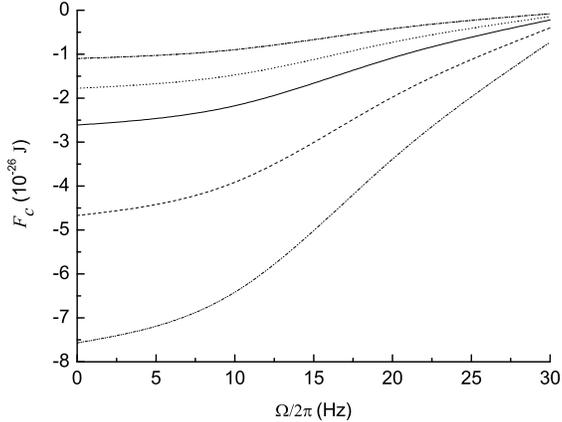, height=2.6in} \caption{The
condensation energy at $T=0.9T_C$ as a function of $\Omega$. The
solid line is the condensation energy with the weak-interaction
expressions of $c_\alpha$ and $c_\beta$.   The dotted-dash and
double-dotted-dashed lines are the condensation energies with
$c_\alpha$ decreased and increased by $20\%$ respectively.  The
dashed and dotted lines are the condensation energies with $c_\beta$
decreased and increased by $20\%$ respectively. } \label{Ecp}
\end{figure}

So far we use the weak-interaction expressions of the parameters
$c_\alpha$ and $c_\beta$ to obtain vortex properties. It is likely
that the accurate expressions of these parameters in the unitary
limit are different.  Here we examine how sensitive the vortex
properties are to these parameters.  In Fig. (\ref{n10p}), the
density profiles of the vortex at $T=0.9T_C$ with various values of
$c_\alpha$ and $c_\beta$ are plotted.
As shown in Fig. (\ref{n10p}), when $c_\alpha$ increases or
$c_\beta$ decreases, the dip at the center of the vortex becomes
bigger.  This is due to the increase of order parameter and the
increase of the absolute value of the condensation energy as shown
in Fig. (\ref{Ecp}).
If the vortex structure can be measured in the experiment, the
values of the parameters $c_\alpha$ and $c_\beta$ can be pinned
down, which is very helpful to obtain other properties of the
system.

In summary, we have developed a phenomenological theory to study the
vortex properties of a superfluid Fermi gas in the unitary region.
This charge-neutral system under rotation is mapped onto an electron
system in a vector potential.  In the framework of the
Ginzburg-Landau theory, the order-parameter and density
distributions in the equilibrium state are solved numerically from
the saddle-point equations of the free energy.  We identify that the
superfluid density is four times the magnitude of the order
parameter squared, which can be determined from the angular-momentum
density.  The size of the vortex core is found to increase with the
rotation frequency due to the density decrease near the core.
Thermodynamical properties of the vortex, such as the free energy
and angular momentum, are also obtained. Our theoretical results can
be directly tested in experiments.

Although the vortex properties in the unitary region are studied
extensively in this paper, more work are needed to study the
rotation effect in other regions of the BEC-BCS crossover, such as
the crossover region and the BEC side.  It is also interesting to
extend this phenomenological approach to study the properties of
vortex lattices in the BEC-BCS crossover.

We would like to thank T.-L. Ho, P. Ao, and D.P. Li for helpful
discussions.  This work is supported by NSFC under Grant No.
90303008.

\end{document}